\documentclass[prl,letterpaper,aps,twocolumn]{revtex4}
\usepackage{graphicx}
\usepackage{amsmath}
\begin{document}
\title{Stability of Superflow for Ultracold Fermions in Optical Lattices}
\author{A. A. Burkov}
\affiliation{Department of Physics and Astronomy, University of Waterloo, 
Waterloo, Ontario, Canada N2L 3G1}
\author{Arun Paramekanti}
\affiliation{Department of Physics, University of Toronto, Toronto, Ontario, 
Canada M5S 1A7}.  
\date{\today}
\begin{abstract}
Motivated by recent observations of superfluidity of ultracold 
fermions in optical lattices, we investigate the stability of superfluid 
flow of paired fermions in the lowest band of a strong optical lattice.
For fillings close to one fermion per site, we show that superflow breaks 
down via a dynamical instability leading to a transient density wave. 
At lower fillings,
there is a distinct dynamical instability, where the superfluid stiffness
becomes negative; this evolves, with increasing pairing interaction,
from the fermion pair breaking instability to the 
well-known dynamical instability of lattice bosons.
Our most interesting finding is the existence of a transition, over
a range of fillings close to one fermion per site, from the fermion 
depairing instability to 
the density wave instability as the strength of the pairing interaction is 
increased. 
\end{abstract}
\maketitle
One of the fundamental nonequilibrium properties of a superfluid is the
critical velocity beyond which superflow breaks down. A closely 
related 
quantity is the critical flow momentum, $Q_c$, defined as the maximum 
sustainable 
phase gradient in the superfluid. This critical momentum conveys
information about important length scales in the superfluid as is
easily seen for dilute quantum gases.
For dilute bosonic superfluids, the Landau criterion tells us that
superfluidity breaks down when the flow velocity exceeds the velocity of the 
Bogoliubov `phonons' of the superfluid. The critical
flow momentum is then easily shown to be the inverse healing length of the 
superfluid. For weakly paired fermionic superfluids the superflow is 
limited by the small pairing gap.
In this BCS regime, fermions depair at a critical flow 
momentum, which is the inverse Cooper pair size. As one tunes the
interaction between fermions in a trapped Fermi gas
through the BCS to BEC crossover,
the critical momentum evolves from the depairing momentum of 
fermions to the inverse healing length of molecular bosons
~\cite{Miller07,Randeria06}. 

A different route to the breakdown of superfluidity is through a 
dynamical instability as has been experimentally observed for
bosonic atoms in optical lattices \cite{dynexp}. 
For weakly correlated bosons in a lattice, there is a 
critical momentum, $Q_c \sim \hbar(\pi/2a)$, where $a$ is the optical 
lattice constant, at which superflow breaks down as the effective
superfluid density becomes negative \cite{Niu01}.
(Henceforth we will set $\hbar\!\!=\!\!1$, $a\!\!=\!\!1$.)
For bosons at commensurate filling this dynamical 
instability has been shown \cite{Altman05,Ketterle07}
to be smoothly connected to the equilibrium superfluid to Mott insulator
transition of bosons \cite{MottIns}, in the sense that the critical 
momentum for the onset of this dynamical instability tends to zero as the 
system is tuned towards the Mott transition.
Close to the Mott transition, $Q_c \sim 1/\xi$,
where $\xi$ is the diverging correlation length for fluctuations near the
superfluid to Mott insulator transition. Thus, the
critical flow momentum in this case conveys nontrivial information about an
equilibrium quantum phase transition.

\begin{figure}[t]
\includegraphics[width=8cm,height=7cm]{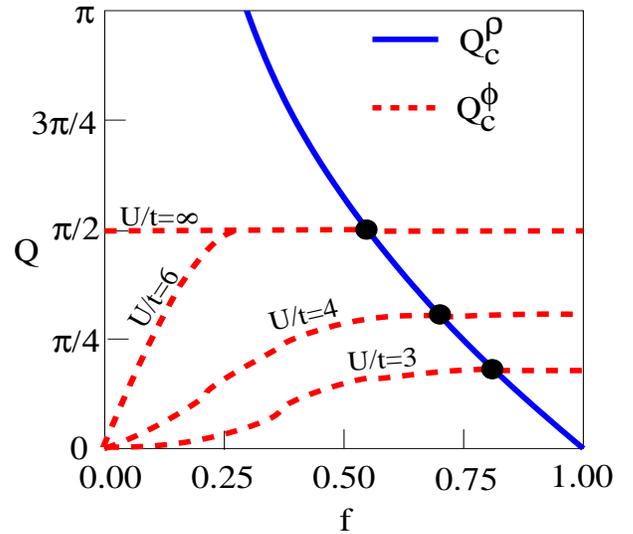}
\caption{(Color online) 
Critical flow momenta for phase stiffness dynamical instability, 
$Q_c^{\phi}$, and density wave dynamical instability, $Q_c^{\rho}$, 
versus fermion filling $f$ at various values of U/t in three 
dimensions~\cite{footnote2}. For $Q^\phi_c \lesssim \pi/2$, the 
fermions depair at $Q \approx Q_c^\phi$, but they depair at larger 
values of $Q$ once $Q_c^\phi=\pi/2$. Black dots indicate points where a sharp transition 
from the phase stiffness to the density wave instability occurs.}
\label{fig:phasediag}
\end{figure}    
In this Letter, we focus on the remarkable and rich physics associated 
with the breakdown of superflow of paired fermions in an optical lattice. 
In the strong lattice potential 
limit, the ultracold fermion system can be described by a single band 
negative-$U$ Hubbard model \cite{Ranninger}, which we
define on a $d$-dimensional hypercubic lattice:
\begin{equation}
\label{eq:1}
H_{\rm HUB} = - \sum_{i,j} t_{ij} c^{\dag}_{i \sigma} 
c^{\vphantom \dag}_{j \sigma} - \mu \sum_i n_i -
U \sum_i n_{i \uparrow} n_{i \downarrow}, 
\end{equation}
where $c^{\dag}_{i \sigma}$ creates a fermionic atom in the hyperfine spin 
state $\sigma$ on a lattice site $i$. This model is well known to have a
superfluid ground state. The phase diagram in Fig.~1 summarizes the two
main new results of our work.
(i) For fillings far from one fermion per site ($f\!=\!1$), 
we show that there is a dynamical instability at a critical momentum
$Q_c^\phi$ where the effective superfluid 
density becomes negative leading to the breakdown of superflow. For large
$U/t$, this is related to the dynamical instability of lattice bosons, 
discussed earlier, while for small $U/t$ it is closely related to depairing 
of fermions.
(ii) We show that for $f\!=\!1$, there is an insulating crystal state which
is degenerate (or nearly degenerate) with the uniform superfluid. For a 
range of fillings around $f\!=\!1$, we show that this competing crystal state 
leads to the breakdown of superflow
via a {\it distinct} dynamical instability at $Q_c^\rho$, 
which leads to a transient {\it density wave} as shown in Fig.~2. 
The critical momentum for this instability is the inverse density 
correlation length (at the relevant wavevector) in the uniform superfluid.

To observe this dynamical instability in cold atom experiments, 
two ingredients are necessary. First, one needs counterpropagating
laser beams, which are frequency-detuned from each other to generate 
a moving optical lattice. Indeed, a recent experiment using such 
moving lattices has successfully 
measured the critical current of paired fermions \cite{Miller07} 
in a weak optical lattice. Second, one needs to be able to detect the
density wave order, induced by the dynamical instability. This could be 
done using light scattering or noise \cite{noise} measurements. Light
scattering at intermediate timescales should be able to detect broadened 
Bragg spots associated with the emergent crystalline order.

While our work is in part inspired by recent experiments demonstrating 
fermionic superfluidity 
in an optical lattice \cite{Chin06}, a broader motivation is 
to explore how {\em competing broken symmetry states}, such as charge
density waves, lead to a breakdown of superflow in uniform strongly 
correlated superfluids --- this issue is of great importance for systems 
ranging from high temperature superconductors \cite{Tranquada} to excitonic 
superfluids in quantum Hall bilayers \cite{MacDonald03}.

We begin with the observation \cite{SO4} that if the
fermions only hop between nearest neighbor sites and if the density is
fixed to be one fermion per site ($f\!=\!1$), the Hubbard model 
Eq.(\ref{eq:1}) possesses an extra pseudospin SU(2) symmetry in addition 
to the usual SU(2) spin rotation symmetry. Let us define the local
version of Anderson's pseudospin operators \cite{Anderson58} as
$T_i^+ = c^{\dag}_{i\uparrow} c^{\dag}_{i \downarrow},\,\, 
T_i^- = c^{\vphantom\dag}_{i\downarrow} c^{\vphantom\dag}_{i \uparrow},\,\, 
T^z_i = \frac{1}{2} (\sum_\sigma c^{\dag}_{i \sigma} 
c^{\vphantom \dag}_{i \sigma} - 1)$.
One can show that the total pseudospin operators
$\sum_i s_i T_i^\pm, \sum_i T^z_i$ commute with the negative-$U$
Hubbard Hamiltonian, where
$s_i=+1,-1$ on the two sublattices of the hypercubic lattice.
This leads to an exact degeneracy 
between the superfluid and the ``checkerboard'' $(\pi,\ldots,\pi)$ charge 
density wave (CDW) states at $f\!=\!1$. In pseudospin language, the
superfluid corresponds to ferromagnetic 
ordering of $T^{\pm}$ while the CDW corresponds to antiferromagnetic 
ordering of $T^z$.

This degeneracy, and what lifts it, is most clearly revealed in the 
large $U/t$ limit of Eq.(\ref{eq:1}), which can be realized in cold atom 
experiments by going to a stronger optical lattice potential. In this limit, 
we can view the presence (absence) of a fermion pair at a site as a 
pseudospin up (down) configuration, and the Hamiltonian in Eq.(\ref{eq:1}) 
maps onto a spin-a-half pseudospin model:
\begin{equation}
\label{eq:2}
H_{\rm xxz}\!\! =\!\! \sum_{i,j} \frac{J_{ij}}{2} \left[T^z_i T^z_j\! -\! \frac{1}{2} \left(T^+_i 
T^-_j\!\! +\!\!h.c. \right) \right]\!\! -\!\! 2 \mu \sum_i T^z_i, 
\end{equation}
where $J_{ij}=4 t^2_{ij} /U$.  
At $\mu = 0$, which corresponds 
to $f\!=\!2 \langle T^z_i \rangle +1\! =\! 1$, and assuming $J_{ij}$ connects 
only neighboring sites with a strength $J \equiv 4 t^2/U$, Eq.(\ref{eq:2}) 
has SU(2) pseudospin symmetry. 

This pseudospin SU(2) symmetry is broken in favor of the superfluid
for nonzero $\mu$, which dopes the system away from $f\!=\!1$. It is lifted 
even 
when $\mu=0$ if the pseudospin exchange $J_{ij}$ connects next neighbor sites, 
with a magnitude $J'\equiv 4 t'^2/U \ll J$, in which case the `classical'
ferromagnetic state of $T^{\pm}$ has a lower energy density,
compared to the antiferromagnetic state of $T^z$, by an amount proportional
to $J'$. 
The question we would like to address here is how the presence of this
broken symmetry crystal as a competing ordered state close by in energy
leads to an instability of the uniform flowing superfluid. We begin 
with a large $U/t$ analysis, where the
physics is fully described by our pseudospin model Eq.(\ref{eq:2}); 
we then generalize our results to all values of $U/t$. For simplicity we
set $t'=0$, the $t'\neq 0$ case is not substantially different.

The evolution of the metastable current-carrying state of 
(\ref{eq:3}) as a function of the filling $f$ and the flow momentum $Q$
can be conveniently analyzed in the semiclassical limit of large
pseudospin. Let $H_{c\ell}$ denote
the classical Hamiltonian, obtained by replacing the pseudospin 
operators in $H_{\rm xxz}$
by angular momentum vectors $\vec{T}_i \equiv T \vec{\Omega}_i$,
where $\vec{\Omega}_i=(\cos\theta_i,\sin\theta_i\cos\phi_i,\sin\theta_i
\sin\phi_i)$ are unit vectors and $T$ denotes the magnitude
of the angular momentum ($T=1/2$ in our case). With this parametrization,
\begin{eqnarray}
\label{eq:3}
\!H_{c\ell}[\Omega]\!\!&\!\!=\!\!&
\!\!\frac{T^2}{2} \sum_{ij} J_{ij} \left[ \cos\theta_i \cos\theta_j
\!\!-\!\!\sin\theta_i \sin\theta_j \cos(\phi_i\!\! -\!\! \phi_j)\right] 
\nonumber\\ 
&-& 2 \mu T \sum_i \cos(\theta_i). 
\end{eqnarray}
For brevity, we present analytical results for $f \leq 1$; the physics 
is identical for $f > 1$ when $J'=0$.

Assuming the current flows in the $x$-direction, we begin by extremizing
the classical energy to find $\theta_i = \theta = \arctan[\sqrt{f (2-f)}/(f-1)],
\,\, \phi_i = Q x_i, \,\, \mu = d J T (f - 1) (1 + \gamma_{\bf Q})$,
where $\gamma_{\bf k} = \frac{1}{d}\sum_{i=1,\ldots, d}\cos(k_i)$ and 
we have assumed that the filling $f$, rather than the chemical potential, 
is fixed. 
Stability of the superflow depends on whether this state is a local 
minimum of the energy functional or just a saddle point. 
To answer this question we expand $H_{c\ell}$ to second order in 
fluctuations about the above solution and Fourier transform it to
obtain:
\begin{eqnarray}
\label{eq:4}
H_{c\ell}[\Omega]&=&\sum_{\bf k}\left[ \rho_{\theta \theta} ({\bf k}) \delta \theta^*_{\bf k} \delta \theta^{\vphantom *}_{\bf k} + 
\rho_{\phi \phi}({\bf k}) \delta \phi^*_{\bf k} \delta \phi^{\vphantom *}_{\bf k} \right. \nonumber \\
&+&\left. \rho_{\theta \phi}({\bf k}) 
\delta \theta^*_{\bf k} \delta \phi^{\vphantom *}_{\bf k} + 
\rho_{\theta \phi}^*({\bf k}) \delta \phi^*_{\bf k} \delta \theta^{\vphantom *}_{\bf k}\right],
\end{eqnarray}
where we have subtracted the energy of the uniform flowing state.
Defining $\eta_{\bf Q}({\bf k})\equiv (\gamma_{{\bf k} +{\bf Q}} + 
\gamma_{{\bf k} - {\bf Q}})/2$, the three different 
stiffness coefficients 
in (\ref{eq:4}) are given by:
\begin{eqnarray}
\label{eq:5}
\!\!\!\rho_{\theta \theta}({\bf k})\!\!&\!=\!&\!\! \frac{\partial^2 H_{c\ell}}{\partial \theta^*_{\bf k} \partial \theta^{\vphantom *}_{\bf k}}\!\! =\!\! 
d J T^2 \left[ \gamma_{\bf Q}\!\! - \eta_{\bf Q}({\bf k}) \right. 
\nonumber\\
&+& \left.\left(\gamma_{\bf k} + \eta_{\bf Q}({\bf k})\right) 
\sin^2(\theta)\right], \nonumber \\
\rho_{\phi \phi}({\bf k})\!&\!=\!&\!\frac{\partial^2 H_{c\ell}}{\partial \phi^*_{\bf k} \partial \phi^{\vphantom *}_{\bf k}} =
d J T^2 \sin^2(\theta) \left(\gamma_{\bf Q} - \eta_{\bf Q}({\bf k}) \right), 
\nonumber \\
\rho_{\theta \phi}({\bf k})\!&\!=\!&\!\frac{\partial^2 H_{c\ell}}{\partial \theta^*_{\bf k} \partial \phi^{\vphantom *}_{\bf k}}\!\! =
\!\! - i \frac{d J T^2 
\sin(2 \theta)}{4} (\gamma_{{\bf k} + {\bf Q}}\!\!  -\!\!  \gamma_{{\bf k} - {\bf Q}}). 
\end{eqnarray}
Here $\rho_{\theta \theta}({\bf k})$ is the inverse density
susceptibility at wavevector ${\bf k}$, $\rho_{\phi \phi}({\bf k})$ is 
the phase stiffness and $\rho_{\theta \phi}({\bf k})$ is related to the 
current, transported by the mode ${\bf k}$.  

The flowing state is a stable local minimum provided the 
following conditions are satisfied:
\begin{equation}
\label{eq:6}
\rho_{\theta \theta}({\bf k}), \,\, \rho_{\phi \phi}({\bf k}) > 0, \,\, 
\rho_{\theta \theta}({\bf k})  \rho_{\phi \phi}({\bf k}) > |\rho_{\theta \phi}({\bf k})|^2, \,\, \forall {\bf k}. 
\end{equation}
There are thus three distinct instabilities, which can lead to the decay of 
superflow. 
For reasons that will become clear below, the instabilities corresponding to 
vanishing 
$\rho_{\theta \theta}$ or $\rho_{\phi \phi}$ are called dynamical,
and we will refer to the critical flow momenta, where these occur, as
$Q^{\rho}_c$ and $Q^\phi_c$ respectively, while
instability corresponding to the violation of the 
last condition in Eq.(\ref{eq:6}) is called Landau or energetic 
instability \cite{Niu01} and we will refer to the critical flow momentum
for this instability as $Q_c^{Lan}$.

Both dynamical and Landau instabilities have the same physical origin, 
namely the superflow-carrying state becoming a saddle-point 
of the energy functional instead of a local minimum.
In order to understand the difference between these instabilities, we turn 
to an analysis of the dynamics. Defining new variables $p_{\bf k} = 
T \sin(\theta) \delta \theta_{\bf k} = p_{1 \bf k} + i p_{2 \bf k}$ and 
$q_{\bf k} = \delta \phi_{\bf k}= q_{1 \bf k} + i q_{2 {\bf k}}$,
it is easy to show that $p_{\nu \bf k}$ and $q_{\nu \bf k}$ are canonically 
conjugate variables which thus obey Hamiltonian quasiclassical equations 
of motion. Rewritten in terms of $p_{\nu}$ and $q_{\nu}$, the 
fluctuation Hamiltonian Eq.(\ref{eq:4}) has the form of a Hamiltonian of a set of decoupled harmonic oscillators in two dimensions, labelled by ${\bf k}$, with characteristic frequency 
$\omega_{0 \bf k} = 2\sqrt{\rho_{\theta \theta}({\bf k}) \rho_{\phi \phi}({\bf k})}/ T \sin(\theta)$, 
rotating with frequency $\Omega_{\bf k} = 2 \,|\rho_{\theta \phi}({\bf k})|/ T \sin(\theta)$. 
In this language the two dynamical instabilities correspond to the point
where the frequency of one or more of the oscillators becomes imaginary, while Landau instability corresponds to the rotation frequency 
exceeding the oscillator frequency.
Diagonalizing the  Hamiltonian by a canonical transformation 
we obtain the eigenfrequencies $\omega_{\bf k} = 
\omega_{0\bf k} \pm \Omega_{\bf k}$. {\em Spectral stability}
requires $\rho_{\theta \theta}({\bf k})>0, \rho_{\phi \phi}({\bf k}) > 0$. 
When either one of these become negative, the eigenmode 
frequencies acquire imaginary parts, which means that isoenergetic 
orbits in the phase space become open, i.e. the motion in the neighborhood 
of the $p_{\bf k} = 0, q_{\bf k} = 0$ saddle point becomes dynamically 
unstable. The motion in the Landau-unstable regime is, however,  
{\em dynamically stable} in this linearized problem, 
since isoenergetic orbits in this case are closed.
Once mode-coupling, due to nonlinearities or external 
potentials, is included the motion will most likely immediately 
become unstable.
We however expect that at low temperatures and in two and three dimensions, 
where quantum corrections to the quasiclassical dynamics are weak 
\cite{footnote1}, the Landau instability in our system
system will develop on much longer time scales compared to dynamical 
instabilities. 

After this preliminary discussion of the instabilities, we now calculate the 
dynamical phase diagram of our system as a function of the flow 
momentum $Q$ and the filling $f$. 
Straightforward analysis of Eq.(\ref{eq:5}) gives the following results
at $T\!\!=\!\!1/2$:
\begin{eqnarray}
\label{eq:7}
\cos Q_c^{\phi}(f)&=& 0, \nonumber \\
\cos Q_c^{\rho}(f) &=& 
\frac{1 - (2 d - 1) (1 - f)^2}{1 + (1 - f)^2}, \nonumber \\
\cos Q_c^{Lan}(f)&=&\frac{A(f,d)
- (2 d - 1) f (2 - f)}{2 [1+ (1-f)^2]},
\end{eqnarray} 
where 
\begin{equation}
\!\!\!\!A(f,d)\!\equiv\!\sqrt{(2 d-1)^2 f^2 (2-f)^2\!\! +\!\! 
8 (1-f)^2 [1\!\! +\!\! (1-f)^2]}. \nonumber
\end{equation} 
Focussing on the dynamical instabilities,
the most interesting feature of Eq.(\ref{eq:7}) is singular behavior 
of the critical momenta as functions of the fermion density. Defining
$f^* \equiv 1-1/\sqrt{2 d -1}$, we find that for $f < f^*$, $Q_c^\phi
< Q_c^\rho$, while for $f > f^*$, $Q_c^\rho < Q_c^\phi$.
There is thus a sharp transition, as a function of filling at $f^*$,
from an instability due to a 
vanishing phase stiffness $\rho_{\phi \phi}({\bf k})$ 
for $0 < f < f^*$ to an instability due to a vanishing inverse
susceptibility $\rho_{\theta \theta}({\bf k})$ for $f^* < f < 1$. 
In $d=1$, we find $f^*=0$.
The phase-stiffness-related instability involves all wavevectors of the 
form ${\bf k} = (k_x,0,\ldots,0)$ (for flow in the $x$-direction).
The density modulation dynamical instability, by contrast, occurs at a 
single wavevector $(\pi,\ldots,\pi)$, since this instability is directly 
related to the checkerboard CDW state.  For $d=3$, this CDW related
dynamical instability only exists for $1-1/\sqrt{2} < f < 1$.

An interesting question is what is the state, towards which the uniform flowing 
superfluid becomes unstable, when $Q_c^{\rho}(f) < Q < Q_c^\phi(f)$.
Since the instability is characterized by a divergent density susceptibility 
at $(\pi,\ldots,\pi)$, one might expect 
the resulting state to be a ``flowing supersolid''  \cite{erhai-ss}, in which the 
CDW coexists with superfluid flow.
Indeed, one finds that such a state extremizes the classical energy and has 
lower energy than the uniform state for $Q > Q_c^{\rho}$.   
Analyzing the stability of such a state, however, we find that it is in turn
dynamically unstable due to a negative superfluid stiffness. 

Since there does not appear to be a simple stable equilibrium state
beyond the $(\pi,\ldots,\pi)$ dynamical instability, we numerically
integrate the semiclassical Landau-Lifshitz equations, $\frac{d
\vec{T}_i}{d t} = \{H_{c\ell},\vec{T}_i\} + \eta \vec{T}_i \times
\{H_{c\ell},\vec{T}_i\}$,
where $\{\,\}$ denotes the Poisson bracket, in order to understand
the fate of the system for $Q > Q^\rho_c$. We
assume a phenomenological damping term, $\eta \ll 1$, to mimic weak 
dissipation arising from thermal excitations, which are relevant for 
experiment but not taken into account in our approach. The density
order, expected to emerge beyond $Q^\rho_c$, can be characterized by
the density structure factor $S(G,t)=(4/L) \sum_{i,m} T^z_m(t) T^z_{m+i}(t) 
\exp(-i G x_i)$. To illustrate
the nature of the collective many body dynamics associated with the
density modulation instability, we focus on two 
physical observables: the density order parameter, $\rho_G(t) \equiv
\sqrt{S(G,t)/L}$, at the wavevector $G$ where $S(G,t)$ is the largest
at that time, and the fermion pair current $I_s(t) = (4/L) \sum_i
[T^x_i(t) T^y_{i+1}(t)- T^y_i(t) T^x_{i+1}(t)]$.
We observe (see Fig.~2) that starting from the uniform 
superfluid,
there is an initial growth of a density modulation order parameter, with
a growth rate proportional to $Q-Q^\rho_c$ and independent of
the weak dissipation. The density modulation 
wavevector $G$ fluctuates around $\pi$, which is the ordering wavevector of 
the CDW crystal. The growth of this CDW order leads to a drop in the flow 
momentum as the current decays via phase slips in regions where the local 
density reaches $f\!=\!1$, giving rise to a transient insulating CDW 
patch. At these intermediate timescales, the CDW order is unstable as
discussed and exhibits fluctuations over time scales $~1/J$ (arising from
excited collective modes) as seen in Fig.~1. In the presence of 
dissipation, the system eventually settles down into a new (lower energy) 
steady state which, typically, is a uniformly flowing state, where the CDW 
order has decayed away and the final flow momentum is $Q < Q^\rho_c$. The 
main message from our numerics is that even when the ``flowing supersolid'' 
state is dynamically unstable, a transient density wave 
nevertheless emerges when $Q > Q^\rho_c$. For modest dissipation, its 
fluctuations persist over a time scale set by the 
inverse dissipation rate as seen from the Fig.~2.

\begin{figure}[t]
\includegraphics[width=8cm]{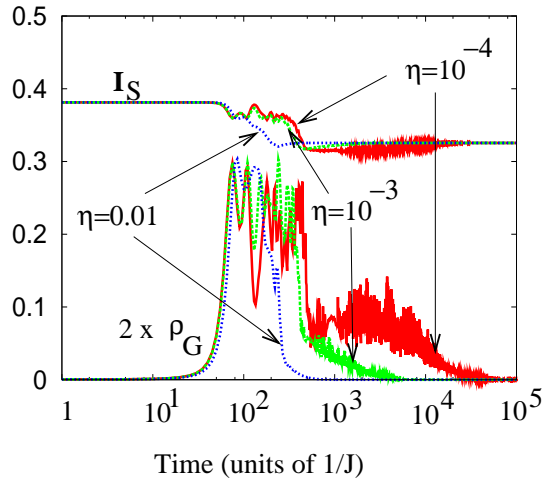}
\caption{(Color online)
Evolution of the current, $I_s$, and the scaled CDW order parameter,
$2 \times \rho_G$, for a one dimensional lattice of 200 sites with 
$f\!=\!0.8$ (corresponding to $Q^\rho_c \approx 0.125 \pi$), 
$Q=0.13\pi$ and indicated values of dissipation.}
\label{fig:LLdynamics}
\end{figure}    

We now generalize the large $U/t$ results, discussed so far, to all values 
of $U/t$.  
For small fillings $f < f^*$ and large $U/t$ we see that the first dynamical 
instability one encounters is at $Q_c^{\phi}=\pi/2$. However, for weaker
couplings, one can view the Anderson pseudospin model as being applicable at
length scales larger than the BCS coherence length, which is proportional
to the fermion pair size $\xi^{pair}$. Thus, this phase stiffness instability 
will occur at $Q_c^{\phi} \sim \pi/2\xi^{pair}$ instead of $\pi/2$, where
$\xi^{pair} \gg 1$ at weak coupling. From this
result it is clear that, at weak coupling, this dynamical instability is
simply related to the fermion depairing instability --- once the flow momentum 
exceeds the inverse fermion pair size, the pairs will break up. Thus, 
when $f < f^*$, there is
a long wavelength dynamical instability associated with vanishing 
$\rho_{\phi\phi}$, which {\it smoothly} evolves from the vicinity of the
pair breaking instability $Q_c^{\phi} \sim 1/\xi^{pair} \ll 1$ when
$U/t \ll 1$,
to $Q^{\phi}_c = \pi/2$ when $U/t \gg 1$.
At fermion densities $f > f^*$ the situation is completely different. 
In this case, for small $U/t$, there is still the above
dynamical instability, which will occur close to 
the depairing instability. However, at larger $U/t$, this instability
is preempted by the distinct density modulation dynamical instability at
$Q^\rho_c$, which is set by the inverse correlation length of the
incipient CDW order in the uniform superfluid. 
The transition between the depairing and the density wave instabilities 
occurs when $Q^\rho_c \, \xi^{pair} \sim 1$, i.e. when the fermion pair size 
becomes smaller than the correlation length of the CDW order.  This 
correlation length diverges at $f\!=\!1$. Since these 
two dynamical instabilities involve completely different wavevectors and 
different susceptibilities, they can not be smoothly connected; thus for
fillings $f^* < f < 1$, there will be a sharp change in the character of 
the dynamical instability with increasing $U/t$. 
This sharp change should 
be contrasted with a smooth evolution of the ground state properties 
with increasing $U/t$ (BCS-BEC crossover). Our result implies that while 
the evolution of the ground state is smooth, the evolution of the excitation 
spectrum in the presence of competing ordered states nearby in energy is not 
smooth, which is revealed in the way the superfluid flow breaks down once 
the critical phase gradient is exceeded. 
We have confirmed the above heuristic arguments by generalizing the BCS 
mean field theory to include flow and density wave order parameter. The 
results of our calculations are summarized in Fig. \ref{fig:phasediag}, 
details will be presented elsewhere \cite{in preparation}.

\acknowledgments{We thank E. Altman, 
E. Demler, A. Griffin, S. Morris, P. Nikolic, T. Sheppard and J. Thywissen 
for discussions. AAB
was supported by a startup grant from the University of Waterloo. AP
acknowledges support from NSERC and the Alfred P. Sloan foundation.}

\end{document}